\documentclass[11pt,a4paper]{article}
\pdfoutput=1

\usepackage{jheppub}
\usepackage{bbm}
\usepackage{amsfonts}
\usepackage{booktabs}
\usepackage{mathrsfs}
\usepackage{dcolumn}
\usepackage{bm}
\usepackage{amsmath}
\usepackage{slashed}
\usepackage{amssymb}
\usepackage{rotating}
\usepackage{epsfig}
\usepackage{graphicx}
\usepackage{subfig}
\usepackage{float}
\usepackage{latexsym}
\usepackage{multirow}
\usepackage{color}
\usepackage{hyperref}
\usepackage{mciteplus}

\newcommand{\SusH}{{\sc SUSY-HIT v1.4}}
\newcommand{\Pyth}{{\sc Pythia 6}}
\newcommand{\DELP}{{\sc Delphes 3}}
\newcommand{\MadG}{{\sc MadGraph}}

\newcommand{\Fastj}{{\sc FastJet-v3.0.6}}
\newcommand{\beq}{\begin{equation}}
\newcommand{\eeq}{\end{equation}}
\newcommand{\chia}{\ensuremath{\tilde \chi^0_1}}
\newcommand{\mchia}{\ensuremath{m_{\tilde\chi^0_1}}}
\newcommand{\chib}{\ensuremath{\tilde\chi^0_2}}
\newcommand{\mchib}{\ensuremath{m_{\tilde\chi^0_2}}}

\newcommand{\mchic}{\ensuremath{m_{\tilde\chi^0_3}}}

\newcommand{\mchid}{\ensuremath{m_{\tilde\chi^0_4}}}
\newcommand{\chga}{\ensuremath{\tilde\chi^\pm_1}}
\newcommand{\mchga}{\ensuremath{m_{\tilde\chi^\pm_1}}}
\newcommand{\chgb}{\ensuremath{\tilde\chi^\pm_2}}

\newcommand{\delm}{\ensuremath{\Delta m}}

\newcommand{\met}{\ensuremath{\slashed E_T}}
\newcommand{\muco}{\ensuremath{\mu_{\rm col}}}

\subheader{
\textrm{\begin{flushright}
APCTP-PRE2015-003 \\
IPMU15-0015
\end{flushright}}}

\title{Accessing the core of naturalness, nearly degenerate \\ higgsinos,
at the LHC}

\author[a]{Chengcheng Han,}
\author[a]{Doyoun Kim,}
\author[a]{Shoaib Munir,}
\author[a,b,c]{and Myeonghun Park}
\affiliation[a]{Asia Pacific Center for Theoretical Physics,\\
San 31, Hyoja-dong, Nam-gu, Pohang 790-784, Republic of Korea}
\affiliation[b]{Department of Physics,\\ Postech, Pohang 790-784, Korea}
\affiliation[c]{Kavli IPMU (WPI),\\ The University of Tokyo, Kashiwa, Chiba 277-8583, Japan}
\emailAdd{hancheng@apctp.org}
\emailAdd{doyoun.kim@apctp.org}
\emailAdd{s.munir@apctp.org}
\emailAdd{parc@apctp.org}

\abstract{The presence of two light higgsinos nearly degenerate in mass is one of the important characteristics of supersymmetric models meeting the naturalness criteria. Probing such higgsinos at the LHC is very challenging, in particular when the mass-splitting between them is less than 5\,GeV. In this study, we analyze such a degenerate higgsino scenario by exploiting the high collinearity between the two muons which originate from the decay of the heavier higgsino into the lighter one and which are accompanied by a high-$p_T$ QCD jet.
Using our method, we can achieve a statistical significance $\sim 2.9\,\sigma$ as well as $S/B \sim 17\%$ with an integrated luminosity of 3000\,fb$^{-1}$ at the 14\,TeV LHC, for the pair production of higgsinos with masses 124\,GeV and 120\,GeV. A good sensitivity can be achieved even for a smaller mass-splitting when the higgsinos are lighter.
}

\arxivnumber{1502.03734}

\begin{document}
\maketitle

\section{\label{intro}Introduction}

One of the key theoretical motivations for low-energy supersymmetry (SUSY) is that it provides a framework in which a light Higgs boson can be obtained without invoking unnatural fine-tuning of theory parameters. However, the Higgs boson discovered recently~\cite{Aad:2012tfa,Chatrchyan:2012ufa} at the Large Hadron Collider (LHC) has a mass around 125\,GeV and signal rates consistent with those predicted by the Standard Model (SM). These properties of the Higgs boson, in conjunction with the non-observation of supersymmetric particles, have resulted in excluding large portions of the parameter space of the Minimal Supersymmetric Standard Model (MSSM) where the naturalness criteria are satisfied.
If the observed Higgs resonance is to be identified with the lightest CP-even Higgs boson, $h$, of the MSSM, TeV-scale SUSY-breaking masses and/or multi-TeV soft trilinear coupling parameters are necessary, so that
the Higgs boson mass can be enhanced sufficiently via radiative corrections~\cite{Carena:2002es,Arbey:2011ab,Carena:2011aa,Cao:2011sn,Cao:2012fz,Cao:2012yn}. Furthermore, null results from gluino searches at the LHC Run-I have pushed the lower limit on its mass to the TeV scale~\cite{CMS:2013cfa,CMS:wwa,TheATLAScollaboration:2013mha,TheATLAScollaboration:2013tha}. While all of this indicates that SUSY lies at the TeV scale, such a heavy sparticle mass spectrum might spoil the naturalness of the MSSM by requiring excessive fine-tuning for generating the correct Higgs boson mass~\cite{Ghilencea:2012gz}.

In the MSSM, the minimization of the tree-level Higgs potential leads to the following relation between the mass of the $Z$ boson, $m_Z$, and the soft SUSY-breaking Higgs sector parameters~\cite{Arnowitt:1992qp}:
\begin{eqnarray}
\frac{M^2_{Z}}{2}=-\mu^{2}+\frac{m^2_{H_d}-m^2_{H_u}\tan^{2}\beta}{\tan^{2}\beta-1}\approx-\mu^{2}-m^2_{H_u}\,.
\label{minimization}
\end{eqnarray}
The last approximation in the above equation assumes $\tan\beta \gtrsim 10$, where $\tan\beta \equiv v_u/v_d$, with $v_u$ being the vacuum expectation value (VeV) of the $u$-type Higgs doublet and $v_d$ that of the $d$-type one. $m_{H_u}$ and $m_{H_d}$ are the soft SUSY-breaking masses of these two Higgs doublets and the parameter $\mu$ is the common mass paramater for the two Higgs superfields, originating in the superpotential of the MSSM. In order to avoid a large fine-tuning in Eq.\,(\ref{minimization}), $\mu$ and $m_{H_u}$ ought to lie in the $\sim 100$\, GeV\ -- 200\,GeV range.

The MSSM contains four neutralinos, $\tilde \chi^0_{1-4}$, which are the mass eigenstates resulting from the mixing of the fermion components of the Higgs superfields, known as the higgsinos ($\widetilde{H}_d^0,\widetilde{H}_u^0$), with those of the gauge superfields, the gauginos ($\widetilde{B}^0,\widetilde{W}^0$). The lightest of these neutralinos is a dark matter (DM) candidate when $R$-parity is conserved. The physical masses of these neutralinos are dependent on the soft SUSY-breaking gaugino mass parameters, $M_{1,2}$, as well as the Higgs-higgsino mass parameter $\mu$ mentioned above. The unification of the soft gaugino masses at some very high scale implies that $M_1$ and $M_2$ are of the same order as the gluino mass parameter $M_{3}$ at the SUSY-breaking scale~\cite{Martin:2009ad,Younkin:2012ui,Akula:2013ioa,Gogoladze:2012yf}. Thus the exclusion limits on the gluino mass from the LHC, together with the requirement of naturalness, lead to a large splitting between the parameters $M_{1,2}$ and $\mu$. This in turn implies small gaugino-higgsino mixing and, after diagonalization of the neutralino mass matrix, two of the physical neutralinos are gaugino-like while the other two are almost purely higgsinos, which are very close to each other in mass. In fact, for $M_1,\, M_2 \gtrsim 1.5$\,TeV and $\mu \sim 150$\,GeV, the mass-splitting, $\mchib - \mchia$, between the two lightest higgsino-like neutralinos is less than 5\,GeV. At the same time, the lighter of the two charginos, \chga, is also a pure higgsino while the heavier, \chgb, a gaugino. In such a scenario, the mass splitting between the lightest chargino and the lightest neutralino, $\mchga - \mchia$, is typically about half of $\mchib - \mchia$.

To search for SUSY in the parameter space regions of the MSSM with nearly degenerate higgsinos is one of the major challenges for particle colliders.
A lot of emphasis in this regard has been laid on the mono-jet, mono-photon or mono-$Z$ searches at the future
experiments~\cite{Chen:1996ap,Gunion:1999jr,Chen:1999yf,Cao:2009uw,Beltran:2010ww,Giudice:2010wb,Goodman:2010ku,Rajaraman:2011wf,Fox:2011pm,Han:2013usa,Schwaller:2013baa,
ATLAS:2012ky,Chatrchyan:2012me,Khachatryan:2014rra,Baer:2014cua,Brooijmans:2014eja,Anandakrishnan:2014exa,Gori:2014oua}.
However, owing to the very small signal rates as well as the statistical limitations, all these channels are expected to show only percent level yields at the 14\,TeV LHC. Several studies~\cite{Rolbiecki:2012gn,Gori:2013ala,Berggren:2013bua,Han:2013kza,Hikasa:2014yra,Han:2014nba,Barenboim:2014kka,Bramante:2014dza,Han:2014xoa,Martin:2014qra,Han:2014sya,Liu:2014lda,Han:2014aea,
Bramante:2014tba,Barr:2015eva} have suggested
  that the presence of extra leptons may help in improving the sensitivity for these processes. In~\cite{Han:2014kaa,Baer:2014kya} such a compressed higgsino spectrum has been probed by tagging an `opposite sign - same flavor' (OS/SF) lepton pair originating from the decay of a heavy neutralino. However, this method only works well when the higgsino mass-splitting is around 10\,GeV or larger. This is for two main reasons. First, for smaller mass-splitting the two leptons produced are too soft to be tagged efficiently. Second, since these leptons are highly collinear, the signal events are diminished by the requirement to isolate them individually.

Therefore, it is imperative to develop new methods for exploring regions of the MSSM parameter space which are consistent with the naturalness criteria~\cite{Brust:2011tb,Papucci:2011wy,Hall:2011aa,Feng:2012jfa,Cao:2012rz,Baer:2012up,Han:2013poa,Han:2013kga,Kowalska:2013ica} but which may have stayed hidden at the LHC so far. In this article, we discuss a method for probing the compressed higgsino spectrum in which the two highly collimated muons produced in the decays of \chib\ are identified as a single object. We explain the event selection procedure, specific to the kinematics of our signal process, which can be employed to reduce the backgrounds. Using some benchmark MSSM points consistent with such a scenario, we analyze the sensitivity that can be achieved at the 14\,TeV LHC using our method.

The article is organized as follows. In section~\ref{model}, we briefly discuss the model parameter configurations leading to the scenario of our interest.
In section~\ref{method} we explain a tagging method for two soft and collimated muons. In section~\ref{results} we discuss our numerical results in detail.
Finally, we present our conclusions in section~\ref{concl}.

\section{\label{model}Nearly mass-degenerate higgsinos in the MSSM}

\begin{figure}[!tbp]
\begin{center}
\includegraphics[width=6.5cm]{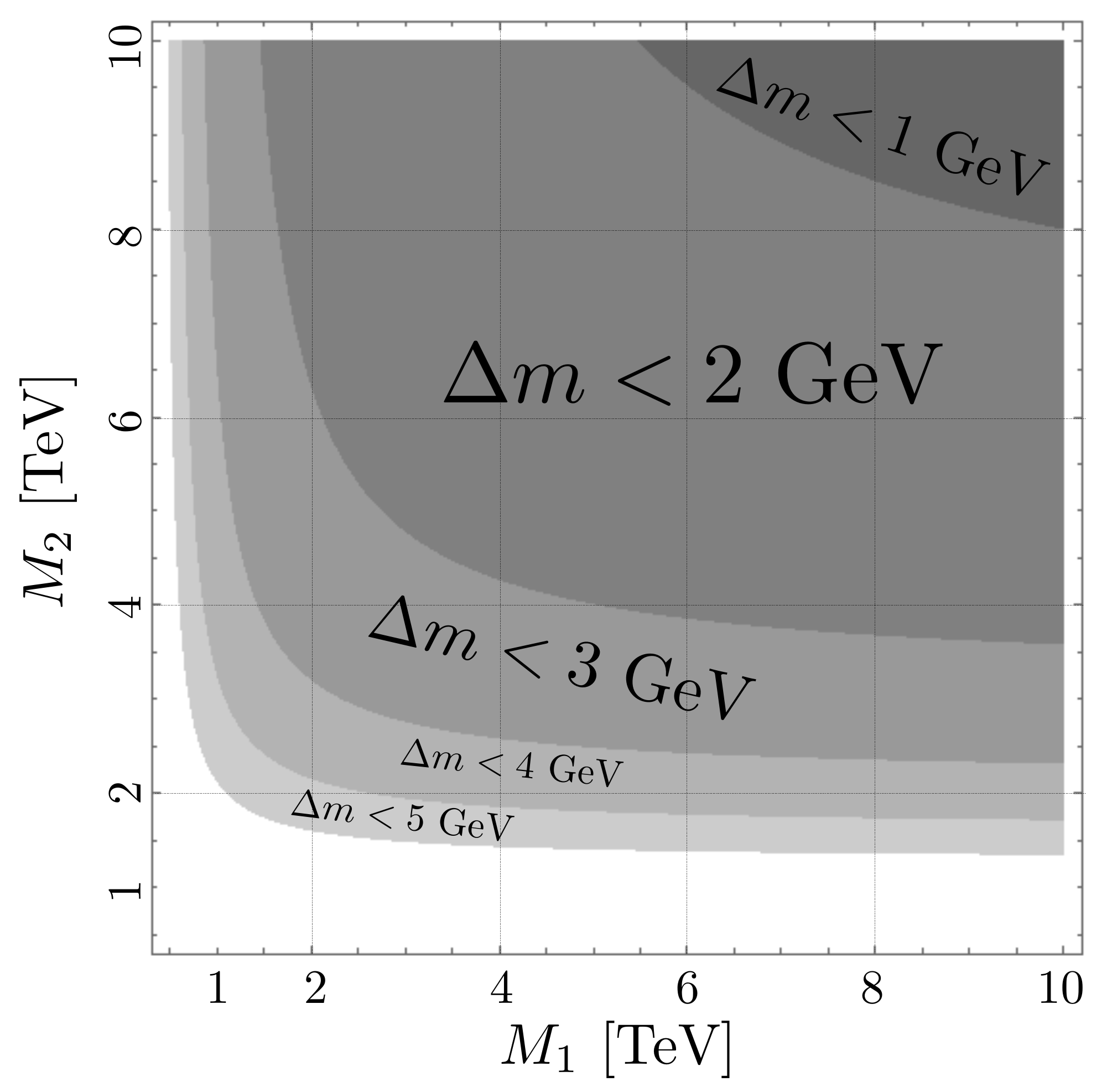}
\caption{Mass-splitting between the two higgsinos as a function of the gaugino mass parameters $M_1$ and $M_2$. $\mu=150$\,GeV and $\tan\beta=30$.}
\label{fig:deltaM}
\end{center}
\end{figure}

The tree-level neutralino mass matrix in the MSSM is written, in the basis $(\widetilde{B}^0,\widetilde{W}^0, \widetilde{H}_d^0,\widetilde{H}^0_u)$, as
\begin{eqnarray}
\begin{array}{c} {\cal M}_{\tilde{\chi}^0} \end{array}
&=& \left( \begin{array}{cccc} M_1 & 0 &-m_W \tan\theta_W \cos\beta &m_W \tan\theta_W \sin\beta  \\ & M_2& m_W \cos\beta &-m_W \sin\beta \\ -m_W \tan\theta_W \cos\beta & m_W \cos\beta & 0 &-\mu \\  m_W \tan\theta_W \sin\beta  &-m_W \sin\beta & -\mu & 0
 \end{array} \right)\,,
\label{eq:neutmass}
\end{eqnarray}
where $m_W$ is the mass of the $W$ boson and $\theta_W$ is the weak mixing angle. The above mass matrix can be diagonalized with an orthogonal real matrix $N$, as
$N M_{\tilde{\chi}^0} N^T= {\rm diag}(\mchia, \mchib, \mchic, \mchid)$, such that $\mchia < \mchib < \mchic < \mchid$. By assuming $M_1, M_2 \gg |\mu|$ in the above mass matrix, one obtains the approximate relation,
\begin{eqnarray}
\delm \equiv \mchib- \mchia  \approx \frac{m_W^2 }{M_2}+ \frac{m_W^2\tan^2\theta_W}{M_1}\,.
\end{eqnarray}
Similarly, using also the chargino mass matrix, one gets $\mchga - \mchia \approx  \frac{\delm}{2}$ (ignoring the terms proportional to $1/\tan\beta$).
The neutralino mass matrix in Eq.~(\ref{eq:neutmass}) is subject to higher order corrections. The diagonalization of the mass matrix in which such corrections have been included (at a certain perturbative order) can be conveniently done numerically using publicly available SUSY mass spectrum calculators. We used the program \SusH\ \cite{Djouadi:2006bz} 
to scan over $M_1$ and $M_2$, both ranging from 0.5\,TeV to 10\,TeV. For this scan we set $\mu=150$\,GeV, $\tan\beta=30$ and the input masses of the SUSY particles other than the electroweakinos to very high values so that they are effectively decoupled. The resultant values of \delm\ are shown in figure~\ref{fig:deltaM}. We see that for $M_{1,2} \gtrsim 1.5$\,TeV the mass-splitting between the higgsino-like \chia\ and \chib\ is always less than 5\,GeV. Also, for such large $M_{1,2}$, the DM direct detection facility XENON1T~\cite{Aprile:2012zx} will not be sensitive to the \chia~\cite{Han:2013usa,Brooijmans:2014eja}, which is the lightest SUSY particle (LSP), when its mass is less than 200\,GeV.

\begin{figure}[tbp]
\begin{center}
\includegraphics[width=8.5cm]{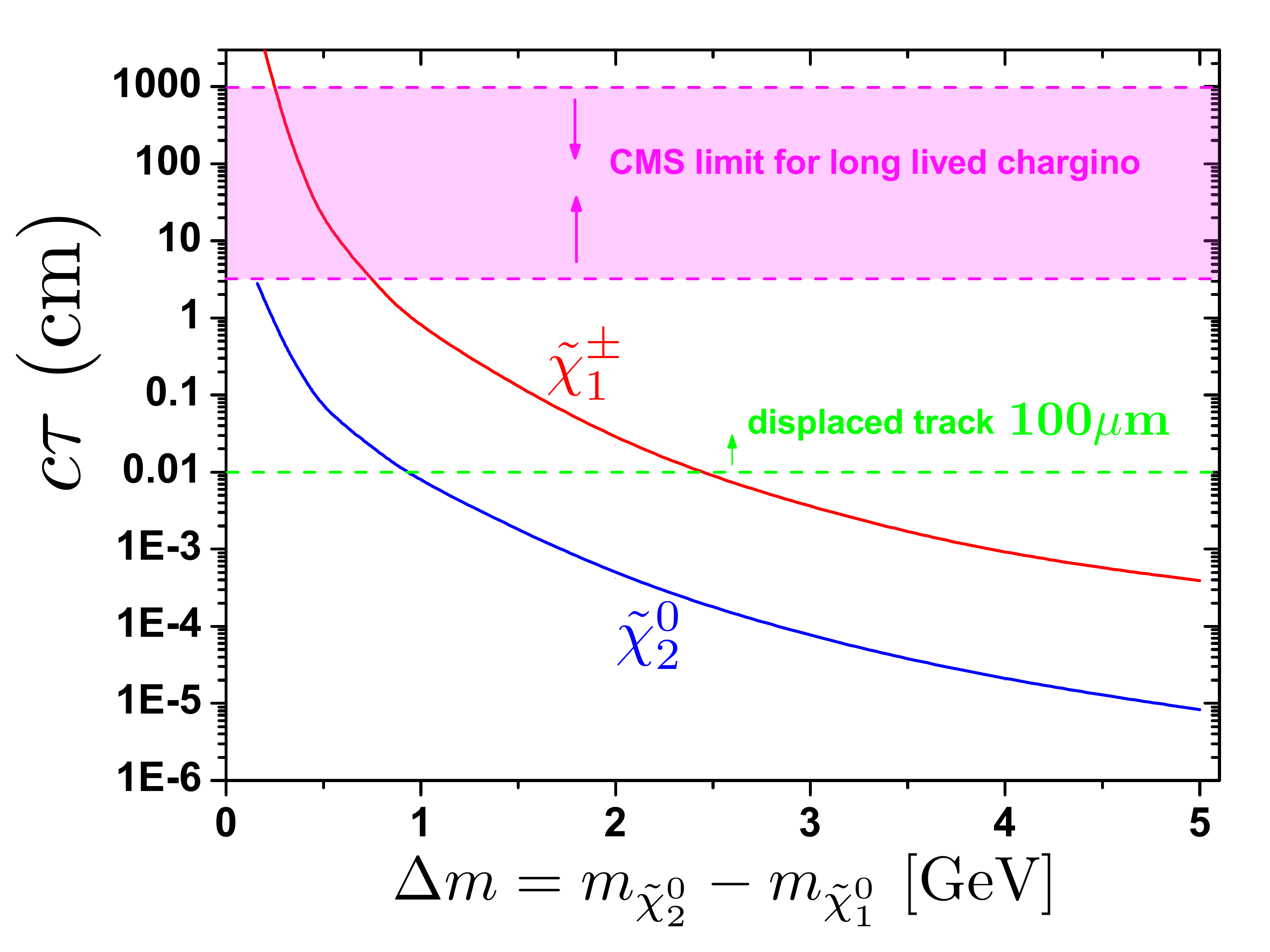}
\caption{Lifetimes of \chga\ and \chib\ as functions of \delm, which is varied by adjusting $M$ ($=M_1=M_2$). The values of $\mu$ and $\tan\beta$ used are the same as in figure\,\ref{fig:deltaM}.}
\label{fig:lifetime}
\end{center}
\end{figure}

Since \chib\ is almost mass-degenerate with \chia, the former can live long enough to leave a secondary vertex in the detector. In figure~\ref{fig:lifetime}, we show the liftime of \chib, calculated with SUSY-HIT, as a function of \delm. We see that, for $\delm < 1$~GeV, the lifetime of \chib\ can be long enough to produce displaced vertices of order 100\,$\mu m$. In fact, for $\delm < 0.1$\,GeV, \chib\ can become collider-stable, so that it leaves the detector before decaying. However, such a tiny higgsino mass-splitting only occurs for $M_{1,2}$ of $\mathcal{O}$(100\,TeV). One also sees in the figure that a chargino has a slightly longer lifetime than a neutralino, which is because $\mchga - \mchia \approx \frac{\delm}{2}$, as noted above. A strong limit on the chargino lifetime, shown by the pink/shaded region in the figure, has recently been obtained by the CMS collaboration~\cite{CMS:2014gxa} for $\delm<1$\,GeV.
At the LHC, a mono-jet along with a displaced vertex larger than 100\,$\mu m$ might help probe the region with $\delm <2.5$\,GeV.

A number of studies, as noted earlier, have explored the MSSM regions with $\delm >10$\,GeV in decays of \chib\ that involve two leptons in the final state. Here we will focus on the splitting region $2 \,\textrm{GeV} \lesssim \delm \lesssim 5$\,GeV for the decay process
\beq
\chib \rightarrow \chia Z^* \rightarrow \chia \ell^+\ell^-\,,
\eeq
with the \chib\ produced via $pp\rightarrow \chib\chia + X$. In our case, due to the small mass-splitting between the higgsinos, the two leptons are generally very soft. Therefore, we only consider muons in the final state on account of  a much cleaner background as well as a much higher trigger efficiency in their case compared to those for taus or electrons.

\begin{figure}[tbp]
\begin{center}
\includegraphics[width=8cm,height=6cm]{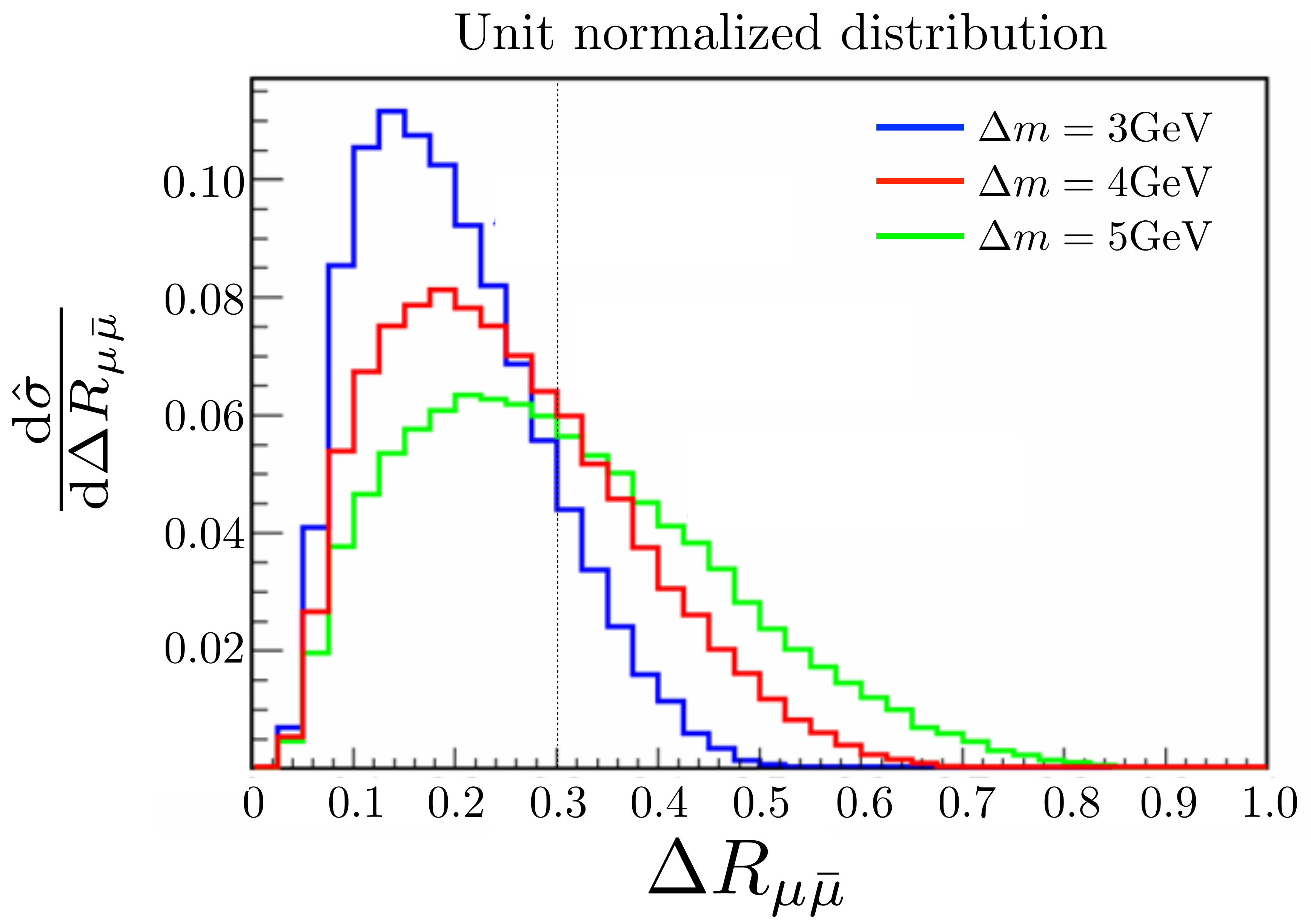}
\caption{The separation $\Delta R_{\mu\bar\mu}$ between the two muons coming from the $Z^*$. The dotted vertical line shows that with the conventional cut, $\Delta R^\textrm{max}=0.3$, the muons in the signal process can not be isolated.}
\label{fig:deltaR}
\end{center}
\end{figure}

In figure~\ref{fig:deltaR} we show the separation, $\Delta R_{\mu\bar\mu} \equiv \sqrt{\Delta\eta^2 + \Delta \phi^2}$ (with $\eta$ being the pseudorapidity and $\phi$ being the azimuthal angle), between the two final-state muons ($p_T(\mu)>$ 5 GeV), for $\delm= 3,4,5$\,GeV. Evidently, the usual isolation criteria for a single lepton, $\Delta R^\textrm{max}=0.3$, will remove a large number of the signal events. Thus we need to use an unconventional reconstruction method for probing such collimated muons and establishing our signal over the SM backgrounds.

\section{\label{method}Reconstructing collimated muons}

In order to probe the highly collinear muons produced for a very small \delm, we cluster them together into one object, $\mu_{\rm col}$, during our simulation of the higgsino pair-production process. This method is similar in concept to the identification  of a `lepton-jet'\ \cite{ArkaniHamed:2008qp,ArkaniHamed:2008qn,Baumgart:2009tn,Katz:2009qq,Cheung:2009su,Falkowski:2010cm} and has already been used recently in analyses of the decays of a light dark photon or of a light scalar or pseudoscalar ($\leq 3$\,GeV)~\cite{Chatrchyan:2012cg} into two or more soft leptons. Instead of imposing the conventional criterion of $\Delta R^\textrm{max}=0.1$ in order to identify the two muons coming from the $Z^*$ as a lepton-jet, we use a modification of the criteria described in the CMS analyses\ \cite{Chatrchyan:2011hr,Chatrchyan:2012cg} for probing collimated muons. Our method is explained below.

\begin{itemize}
\item Capturing \muco: We require $p_T> 5$\,GeV for each muon in the signal, before isolation. In addition to this, we impose the cut $m_{\mu\bar \mu}<5$\,GeV on the invariant mass of the muon pair, since we are only interested in $\Delta m <5$\,GeV.
\item Isolation: To suppress the backgrounds containing muon pairs from meson decays, we apply an isolation criterion, $I_\textrm{sum} < 3 $\,GeV, on \muco. The isolation parameter $I_\textrm{sum}$ is defined as the scalar sum of the transverse momenta of all additional charged tracks, each with $p_T > 0.5$\,GeV, within a cone centered along the momentum vector of \muco\ and satisfying $\Delta R^\textrm{max} = 0.5$.
\end{itemize}

For the Monte Carlo simulations, we generated the parton-level signal and background events with \MadG\_aMC$@$NLO\ \cite{Alwall:2014hca}. These events were then passed on \Pyth\ \cite{Sjostrand:2006za} for hadronization and subsequently to the fast detector simulator \DELP\ \cite{deFavereau:2013fsa} interfaced with \Fastj\ \cite{Cacciari:2011ma} for jet-clustering. In \DELP\ we added a class for
\muco\ identification. The jets were clustered using the anti-$k_T$\ \cite{Cacciari:2008gp} algorithm with $\Delta R^\textrm{max}$ set to 0.4. As a test of the implementation of our method, we first performed simulations for the benchmark points provided in the CMS analysis\ \cite{Chatrchyan:2012cg} and found our results to be within $5\%$ of the ones presented there, in terms of signal efficiencies.

\begin{figure}[tbp]
\begin{center}
\includegraphics[width=8.0cm]{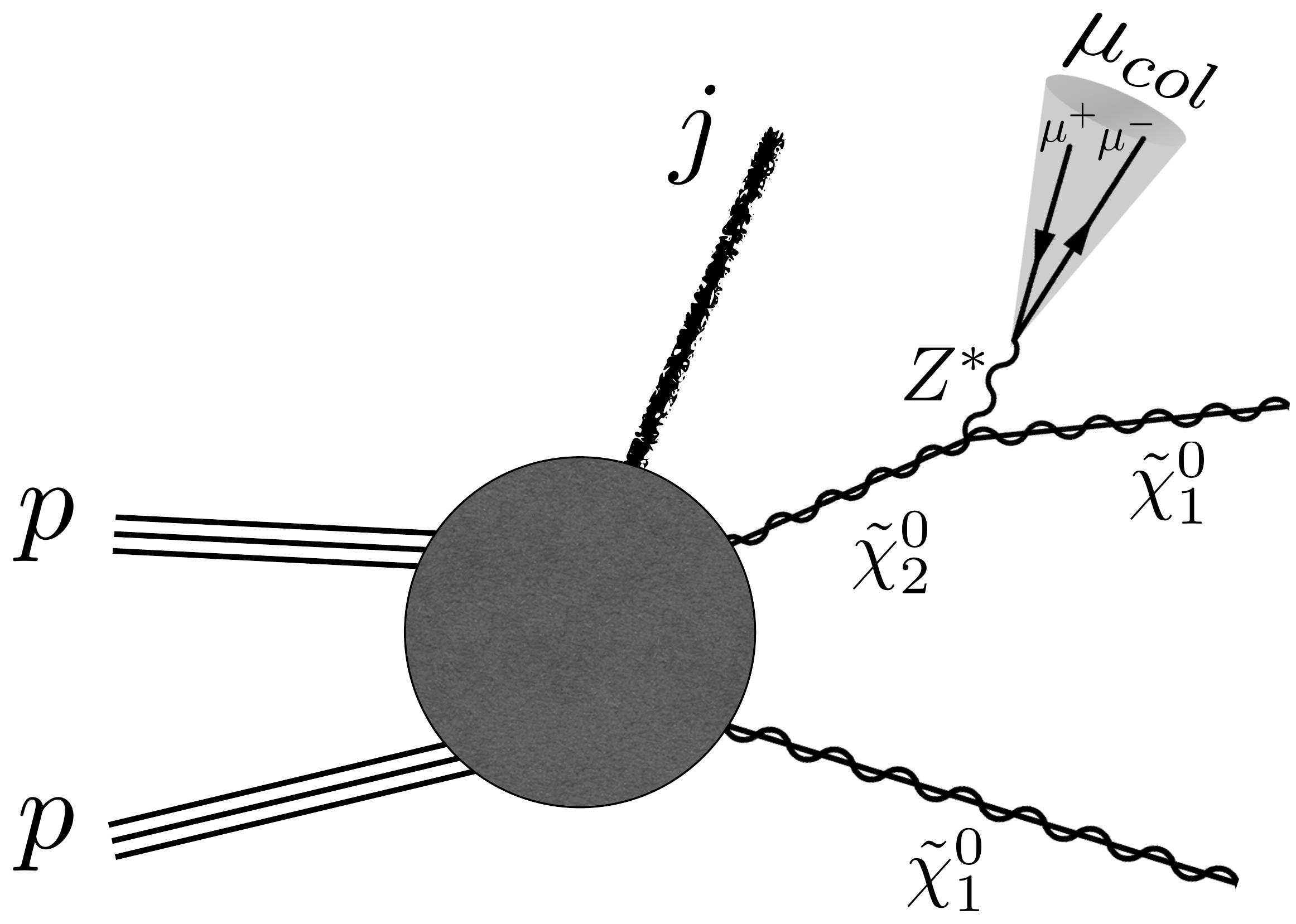}
\caption{Feynman graph for our signal process. The cone signifies a pair of collimated muons $\mu_{\rm col}$ coming from the $Z^*$. $j$ denotes a hard QCD jet from initial state radiation.}
\label{fig:procdiag}
\end{center}
\end{figure}

According to\ \cite{Chatrchyan:2011hr,Chatrchyan:2012cg}, the largest background for our signal process is the $b\bar{b}$ production, which has a cross section of ${\cal O}(10^{8}\,{\rm pb})$. Although requiring $b$-quarks decays, via double semileptonic decays, into pairs of muons (the branching ratio being around 1\%) which are isolated, reduces this background by ${\cal O}(10^{-2})$, it is still huge. We therefore apply a parton-level cut of $p_T > 200$\,GeV on the first leading jet in the signal and the backgrounds. This translates into the requirement of large missing energy, \met, in the final state at the detector level, which almost entirely removes the $b\bar{b}$ background. To include QCD effects \cite{Denner:2012ts}, we use the MLM-scheme to match the additional two jets\ \cite{Caravaglios:1998yr}. We illustrate our signal process in figure~\ref{fig:procdiag}.

\begin{figure}[tbp]
\begin{center}
\includegraphics[width=0.65\textwidth]{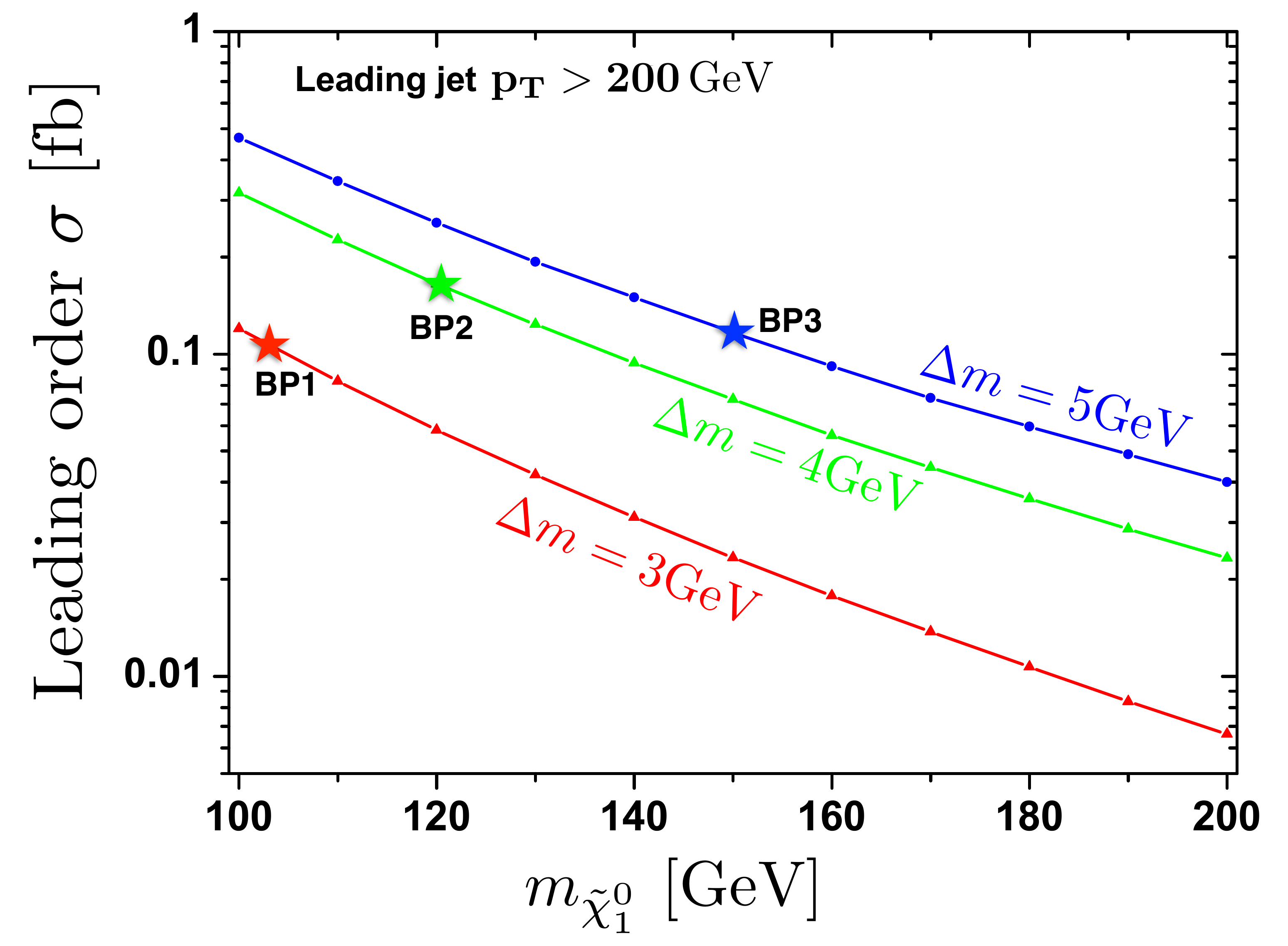}
\caption{The leading order cross sections corresponding to three different values of \delm, as a function of the \chia\ mass. We require $p_T> 200$\,GeV for the leading jet, $p_T>5$\,GeV for each muon and the detector geometry cuts to be satisfied.}
\label{fig:csect}
\end{center}
\end{figure}

In figure~\ref{fig:csect} we show the combined cross section, after imposing the $p_T > 200$\,GeV cut on the leading jet, for our signal process, $pp \rightarrow \chib \chia + X$, and two additional processes, $pp\rightarrow \chib \chga + X$ and $\rightarrow \chib \chib + X$, for three different values of $\delm$. The reason for including the latter two processes is that
the \chga\ in the second process as well as the additional \chib\ in the third process gives very soft products which escape undetected, thus resulting in only \met\ in the final state and thereby mimicking the signal process. We see in the figure that the cross section gets considerably reduced for smaller mass-splitting. However, even with $\delm = 3$\,GeV for a $\sim 200$\,GeV \chia, more than 20 signal events can be obtained at the 14\,TeV LHC with an integrated luminosity, ${\cal L}$, of 3000\,fb$^{-1}$.

For our signal-to-background analysis, we choose three benchmark points (BPs) with a different value of \delm\ each. Some specifics of these points are given in Table~\ref{tab:bench}. The signal cross sections given in the table have been obtained after requiring $p_T> 200$\,GeV for the leading jet and $p_T>5$\,GeV for each muon. The detector geometry cuts on the pseudorapidities of the muons ($|\eta|<2.5$) and the leading jet ($|\eta|<5.0$) have also been imposed.

\begin{table}[th] \centering\caption{Some properties of the three benchmark points analyzed in this study.}
\begin{tabular}{|c|c|c|c|c|c|c|c|}
\hline                                                                         & BP1 & BP2 & BP3 \\
\hline    \mchia\ (GeV)                                            &103.0 & 120.0 & 150.0 \\
\hline    \delm\ (GeV)                                             & 3.0  & 4.0 & 5.0 \\
\hline    Signal cross section (fb)                                       & 0.12 & 0.165 & 0.116 \\
\hline
\end{tabular}
\label{tab:bench}
\end{table}

\subsection{\label{BG}Backgrounds}

After removing the $b\bar b$ background for mono-jet production along with soft muons and \met, the main backgrounds that remain include the following.
\begin{itemize}
\item \underline{$V+\gamma^*+jets$:} A large \met\ results from the $W \rightarrow \ell \nu$ decay or the $Z \rightarrow \nu \nu$ decay and the two collinear
muons originate from the virtual photon, $\gamma^*$. To reduce this background, we require $m_{\mu\bar \mu}>1.0$\,GeV and $\Delta R_{\mu\bar \mu}> 0.1$.

\item \underline{$\tau\bar \tau+jets$:} Each tau decays into a muon and a pair of neutrinos. Due to the large boost in the leading jet, these muons become highly collimated. The neutrinos produced are responsible for a large \met.

\end{itemize}

\noindent We also look into some other backgrounds, described below, which are only ${\cal O}(10^{-2})$ of the above main backgrounds.

\begin{itemize}

\item \underline{$Vb\bar{b}+jets$ and $Zb+jets$:} This background mimics our signal when the $b$-quarks decay into pairs of muons via double semileptonic decays. However, it is suppressed by
a factor $\sim 10^{-4}$ after isolation. We can estimate this background from the ATLAS mono-jet search\ \cite{Aad:2014nra}, where a set of mono-jet cuts is imposed on the backgrounds ($p_T > 280$\,GeV for the leading jet and $\met > 220$\,GeV). The cross section for the $Z(\rightarrow \nu \bar\nu) +jets$ background is around 0.82\,pb. The cross section for $W +jets$, where $W$ decays semileptonically, is around 0.6\,pb. The total cross section for all these backgrounds thus adds up to about 1.4\,pb at the 8\,TeV LHC. At the 14\,TeV LHC, even if one assumes the cross section for these backgrounds to increase by a factor of 10, it will reduce to $\sim 0.3$\,fb after taking into account the suppression of $2\times 10^{-5}$ from the possibility of the $b$-jets giving collinear muons. Thus this background becomes much smaller than the $V+\gamma^*+jets$ backgrounds, which are still around 20\,fb after the mono-jet cuts.

The $W (\rightarrow \mu \nu) b\bar{b}+ jets$ background can also mimic our signal if a $b$-jet is miss-tagged as a muon. The $b$ miss-tagging rate after passing the isolation criteria is less than 0.005, but since this muon tends to have a large separation from a muon resulting from the $W$ decay, this contribution is also small.

\item \underline{$t\bar t+jets$:} After imposing the mono-jet cut and requiring $m_{\mu\bar\mu} < 5$\,GeV, the cross section for this background is reduced to less than 0.1\,fb. We, therefore, do not take it into consideration here.

\end{itemize}

\subsection{Summary of the cuts}
\begin{figure}[tbp]
\begin{center}
\includegraphics[width=0.6\textwidth]{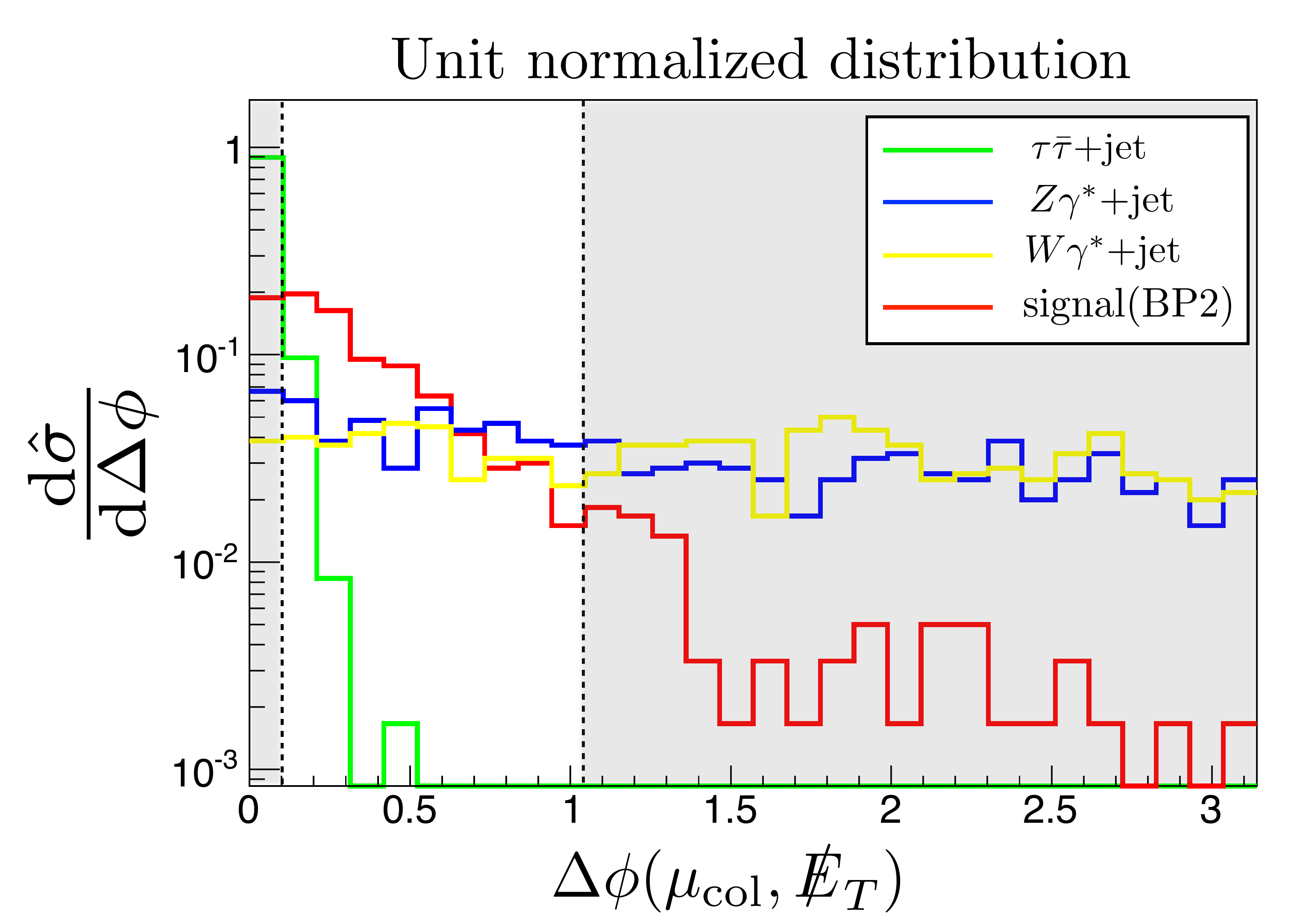}
\caption{The distribution of $\Delta \phi({\muco,\met})$ for the signal and the backgrounds. The signal  corresponds to our BP2. The grey/shaded regions are the ones cut off in our kinematical analysis.}
\label{fig:delphi}
\end{center}
\end{figure}

Below we summarize our cuts based on the discussion above.

\begin{figure}[tbp]
\begin{center}
\includegraphics[width=0.49\textwidth]{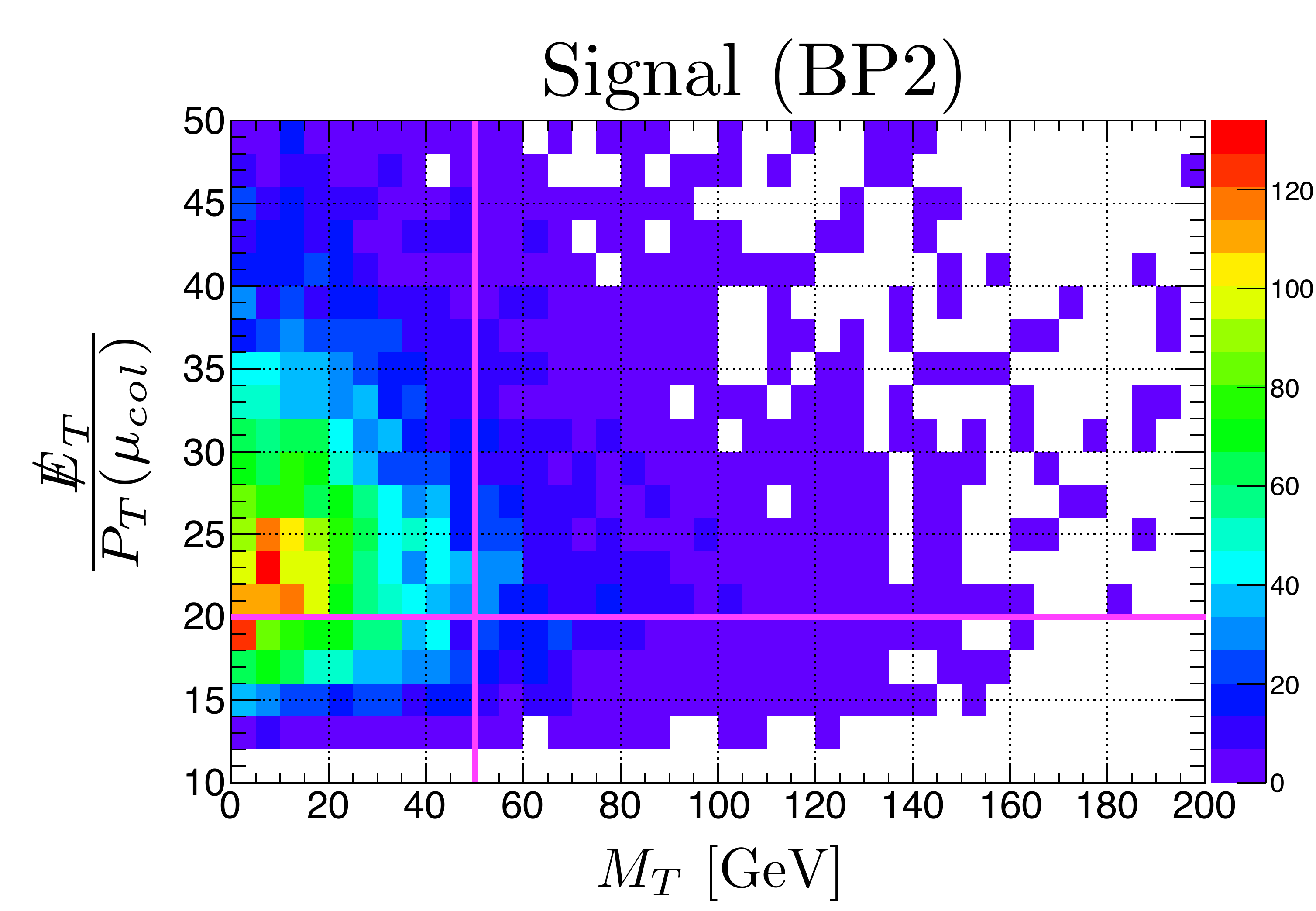}
\includegraphics[width=0.48\textwidth]{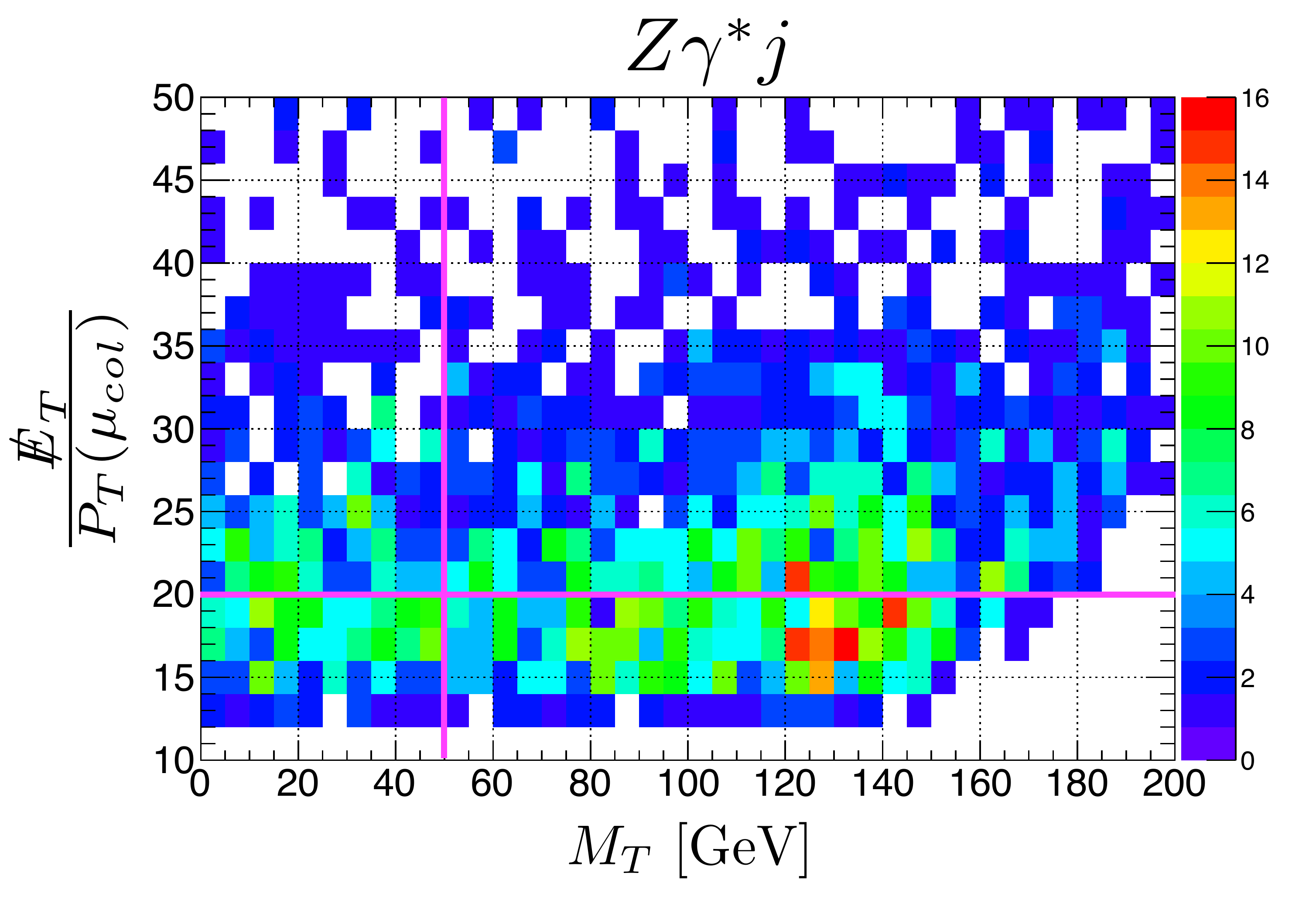}
\caption{$M_T$ vs $\met/p_T(\mu_{\rm col})$ for the signal corresponding to our BP2 (left) and for the $Z\gamma^*$ background (right), after applying the mono-jet cut as well as the basic cuts on $\mu_{\rm col}$. We mark the cuts on $M_T$ and $\met/p_T(\mu_{\rm col})$ with solid magenta lines. Thus the events allowed after the cuts are the ones located in the boxes traced by the lines in the upper-left corners of the figures. The heat map corresponds to the number of events.}
\label{fig:mtvspt}
\end{center}
\end{figure}

\begin{itemize}
\item Mono-jet cut: We require $p_T > 250$\,GeV for the leading jet and veto events which have more than three jets with $p_T > 30$\,GeV. $\Delta \phi$ between the leading jet and the second jet should be larger than 0.4. All the jets are $b$- and $\tau$-vetoed. Any events containing an electron with $p_T$ larger than 10\,GeV are also vetoed. We additionally require $\met > 250$\,GeV and demand exactly one pair of SF/OS muon candidates with each of these muons having $p_T > 5$\,GeV.
\item Basic requirements on $\mu_{\rm col}$: First we define an object $\mu_{\rm col}$ as a two-muon system satisfying $1\,{\rm GeV} < m_{\mu\bar\mu}< 5$\,GeV and $0.1 <\Delta R_{\mu\bar\mu}< 0.5$. Here the cut at the lower end of $\Delta R_{\mu\bar\mu}$ is to remove the main backgrounds where two muons emerge from a $\gamma^*$. The \muco\ is required to be isolated with $I_\textrm{sum} < 3$\,GeV. The $p_T$ of the \muco\ in our signal usually tends to be small. We therefore apply a cut $p_T < 20$\,GeV for the \muco.
\item Cut on $\Delta \phi{(\muco,\met)}$: To further remove the backgrounds, we add a cut on the $\Delta \phi$ between \muco\ and $\met$ as $0.1 <\Delta \phi < \pi /3$. In figure~\ref{fig:delphi} we show the $\Delta \phi{(\muco,\met)}$ distributions for the signal corresponding to our BP2 as well as the backgrounds. We note that imposing the lower cut of $\Delta \phi > 0.1$ removes much of the $\tau\bar \tau+jets$ background and the upper cut leaves only about a third of the $V+\gamma^*+jets$ background. As for the signal, this cut only removes less than 30\% of the events.
\item Mass cut on $\mu_{\rm col}$: To suit the mass of the \muco\ in our signal, we only select events with $1.5\,{\rm GeV} < m_{\mu_{\rm col}} < 4$\,GeV, cutting off also the small window, $3.0\,{\rm GeV}< m_{\mu_{\rm col}} < 3.2$\,GeV, corresponding to the mass of the $J/\Psi$ resonance.
 \item Kinematic cuts: Since in our signal the missing energy originates from two neutralinos, whereas the \met\ in the $V+\gamma^*+jets$ background comes from a single $V$ boson, we expect different transverse mass, $M_T$, distributions of the signal and the backgrounds~\cite{Giudice:2011ib}. In addition, we also impose a cut on $\met/p_T(\muco)$ because the cut on the $p_T$ of the leading jet also results in a boosted \muco. We can use \met\ in this cut instead of the leading jet $p_T$ due to the comparatively much smaller $p_T({\mu_{\rm col}})$. In figure~\ref{fig:mtvspt} we show the distributions of these two variables for the BP2 signal (left) and the $V+\gamma^*+jets$ background (right). We find that an upper cut of $M_T < 50$ GeV and a lower cut of $\met/p_T({\mu_{\rm col}}) > 20$ suppresses the background while allowing most of our signal events.
\end{itemize}

\section{\label{results}Results of the signal-to-background analysis}

In Table~\ref{tab:cutflow} we show the cut-flow for our numerical simulations of the three BPs. We note here that while generating our signal process for each BP we also required $p_T(\mu)> 4$ GeV and $1\,{\rm GeV} <m_{\mu\bar\mu}<5$\,GeV at the parton level. The table shows that the total background cross section after the mono-jet cut is around 23\,fb, which is three orders of magnitude larger than our signal. We observe that the $V + \gamma^*+ jets$ background is much larger in our case than in~\cite{Baer:2014kya}. The reason for the strong suppression of this background in~\cite{Baer:2014kya} is that the two muons coming from the $\gamma^*$ are individually isolated there. However, in our case such an isolation condition will conversely suppress the signal process due to a comparatively smaller mass-splitting between the two higgsinos.

According to Table~\ref{tab:cutflow}, after applying all the cuts, the $V+\gamma^*+ jets$ background is clearly the largest one. The $j\tau\bar \tau$ background only contributes about 10\% to the total since, as noted earlier, the lower cut on $\Delta \phi(\mu_{\rm col},\met)$ reduces this background by nearly an order of magnitude.

\begin{table}[th] \centering\caption{Cut-flow for our signal and background processes. The cross sections are in fb and $S/B$ and $S/\sqrt{B}$ given in the last two rows correspond to ${\cal L} = 3000$\,fb$^{-1}$ at the 14\,TeV LHC.}
\begin{tabular}{|c|c|c|c|c||c|c|c|}
\hline  Cuts           & ~$W\gamma^*j$~ & ~$Z\gamma^*j$~ &~$j\tau\tau$~& Total BKG &BP1  & BP2 & BP3 \\
\hline  Mono-jet                                           & 8.057    & 8.82   & 6.674   &23.0  &0.052        & 0.072    &    0.056       \\
\hline  Basic $\mu_{\rm col}$                                       & 0.753    & 1.05   & 0.314   & 2.1   & 0.041         & 0.042    &    0.028     \\
\hline  $\Delta \phi(\mu_{\rm col},\met)$       & 0.288    & 0.324  & 0.035   &0.65  & 0.028        & 0.030     &   0.020      \\
\hline  $m_{\mu_{\rm col}}$               & 0.106    & 0.118  & 0.024   &0.248 & 0.017       & 0.023      &   0.015      \\
\hline  $M_T\, \&\, \frac{\met}{p_T({\mu_{\rm col}})}$              & 0.037    & 0.044  & 0.011   &0.092  & 0.013       & 0.016      &   0.010      \\
\hline
\multicolumn{5}{|c||}{S/B}                                                                      & 0.14&0.17&0.11 \\
\hline
\multicolumn{5}{|c||}{$S/\sqrt{B}\,(\sigma)$}                                                         &2.4&2.9&1.85 \\
\hline
\end{tabular}
\label{tab:cutflow}
\end{table}

We also note in the table that the highest significance we obtained for the 14\,TeV LHC with ${\cal L} = 3000$\,fb$^{-1}$ is $\sim 3\,\sigma$ and corresponds to our BP2, while $S/B $ for this point is 17\%. We point out here that although both BP1 and BP3 give signal cross sections similar to the one obtained for BP2, the obtained significance is smallest for BP3. The reason is that for BP3 $\delm = 5$\,GeV, so that $p_T(\muco)$ tends to be a little larger than our chosen strong upper cut on $p_T(\muco)$. Another factor is the $m_{\muco}< 4$\,GeV cut, which also removes some signal events in the case of BP3, but not in the case of BP2. Although we can relax the upper cut on $p_T(\muco)$ to around 25\,GeV, it will also result in larger backgrounds and thus the statistical significance for BP3 will not improve.
We should also point out here that for BP1, where $\delm = 3$\,GeV, since $m_{\muco}$ should be less than 3\, GeV, changing the upper cut on $m_{\mu_{col}}$ from 4\,GeV to 3\,GeV would enhance our signal significance. However, in the experimental searches the true value of the higgsino mass-splitting is unknown. We therefore retain the upper cut of 4\,GeV, which suits most of the \delm\ range that we are interested in.

Finally, using our method the minimum \delm\ we have managed to explore is $\sim 3$\,GeV. This is because of the requirement of $p_T > 5$\,GeV for the muons. By imposing a lower cut, $p_T > 4$\,GeV\ \cite{Aad:2012qua}, for the muons, a good sensitivity to even smaller values of \delm\ can be achieved.

\section{\label{concl}Conclusions}

In this article, we have analyzed the possibility of probing natural SUSY scenarios with a highly compressed higgsino mass spectrum at the 14\,TeV LHC, using the collinearity between the two muons produced in such scenarios. We have found that a statistical significance of up to 3\,$\sigma$ as well as $S/B$ up to 17\% can be obtained for $\delm=\mchib - \mchia=4$\,GeV and a $\sim 120$\,GeV \chia\ with an integrated luminosity of 3000\,fb$^{-1}$ at the LHC. In fact, by using our analysis method but further lowering the cut we imposed on the $p_T$ of muons, MSSM parameter space regions with \delm\ even lower than 3\,GeV can be explored.

\section*{Acknowledgments}
This work is supported by the Korea Ministry of Science, ICT and Future Planning, Gyeongsangbuk-Do and Pohang City for Independent Junior Research Groups at the Asia Pacific Center for Theoretical Physics.
MP is also supported by World Premier International Research Center Initiative (WPI Initiative), MEXT, Japan. CCH thanks the Kavli IPMU for its warm hospitality and appreciates encouragement by Hitoshi Murayama.

\bibliographystyle{utphysmcite}
\bibliography{Higgsino_refs}

\end{document}